# Students' Perceived Roles, Opportunities, and Challenges of

# a Generative AI-powered Teachable Agent: A Case of Middle School Math Class

Yukyeong Song (0000-0002-4084-2734) [a]

Jinhee Kim (0000-0002-3365-7354) [b*]

Zifeng Liu (0009-0005-5833-2141) [a]

Chenglu Li (0000-0002-1782-0457) [c]

Wanli Xing (0000-0002-1446-889X) [a]

[a] *School of Teaching and Learning, College of Education, University of Florida, FL, United States*

[b] *Department of STEM Education and Professional Studies, Old Dominion University, Norfolk, VA, United States*

[c] *Educational Psychology, College of Education, The University of Utah, Salt Lake City, UT, United States*

* Corresponding Author Info:

Jinhee Kim

4301 Hampton Blvd, Room 2333, Norfolk, VA 23529

Tel: +1 757 683 5163

email: jhkim@odu.edu

**Abstract**

Ongoing advancements in Generative AI (GenAI) have boosted the potential of applying long-standing "learning-by-teaching" practices in the form of a teachable agent (TA). Despite the recognized roles and opportunities of TAs, less is known about how GenAI could create synergy or introduce challenges in TAs and how students perceived the application of GenAI in TAs. This study explored middle school students' perceived roles, benefits, and challenges of GenAI-powered TAs in an authentic mathematics classroom. Through classroom observation, focus-group interviews, and open-ended surveys of 108 sixth-grade students, we found that students expected the GenAI-powered TA to serve as a learning companion, facilitator, and collaborative problem-solver. Students also expressed the benefits and challenges of GenAI-powered TAs. This study provides implications for the design of educational AI and AI-assisted instruction.

# I. Introduction

The emergence of generative artificial intelligence (GenAI) technologies has motivated educators to appropriate them for educational activities. Among those efforts are teachable agents (TAs). TA is a type of pedagogical agent that instantiates the pedagogical method of "learning-by-teaching" (Chase et al., 2009). Pedagogical agents often take the role of an experienced tutor (Heffernan & Koedinger, 2002) or a knowledgeable learning companion (Kim, 2007), and are purposely designed to arouse students' proactiveness in learning authentically by promoting self-exploration or self-discovery (Biswas et al., 2005). On the other hand, TA features simulating a less-knowledgeable peer to make the students teach them about learning content, through which the process students can learn, reinforce the conceptions or correct their misconceptions (Zhao et al., 2012). Extensive research has demonstrated the advantages of TAs regarding learning achievements and affective domains. For instance, TAs provide students with an opportunity to reorganize their knowledge structure and gain a deeper understanding of domain knowledge (Fantuzzo et al., 1992), enhance metacognition and self-explanation (Roscoe & Chi, 2007), and promote taking responsibility for learning (Chase et al., 2009). Several studies show that lower-achieving students benefit from interacting with a TA, which can help reduce the achievement gap (Chase et al., 2009).

Although discussions have highlighted the benefits of TAs, many challenges still hinder the design of pedagogically meaningful, proactive interactions between TAs and students. These challenges include how TAs learn from students to personalize their learning experiences; how TAs can facilitate learning to be cognitively stimulating; how to design effective scaffolding to guide students' higher-order thinking; and how students and TAs can establish relationships that promote students' sense of ownership and agency in their learning (Silvervarg et al., 2021). GenAI technologies offer the potential to address these limitations by enabling more dynamic and naturalistic conversations, offering personalized and contextualized responses, and adapting to individual learning styles (Johnson & Lester, 2018). However, despite these potentials, researchers have raised concerns about GenAI that it can undermine students' creativity or originality, especially for younger children (Dai et.al, 2023). In addition, the adoption or acceptance of GenAI-powered learning technologies can largely depend on students' perceptions, such as trust (Amoozadeh et al., 2023).

To address such limitations, researchers have indicated the need to incorporate students' perspectives in designing educational GenAI applications and TAs to comprehensively understand student needs in such learning environments (Rose & Shevlin, 2004). Previous research in this field has investigated students' perceptions of general TAs (Biaswas et al., 2016) or general GenAI technologies, such as ChatGPT rather than a specific application like a GenAI-powered TA (Chan & Hu, 2023). In addition, much research focuses on the higher education context, such as college or graduate school levels (e.g., Chan & Hu, 2023; Chen et al., 2024). In the current situation where educational effects and ethical issues of introducing GenAI in K-12 learning settings are actively discussed, it is essential to explore K-12 students' perceptions of the use of GenAI in their classroom. Therefore, this study aims to examine middle school students' views on learning with GenAI-powered TAs within an authentic mathematics classroom to inform the future design of GenAI as TAs in K-12 and tailor instructional strategies to better support AI-assisted learning. Three research questions (RQ) were set to guide this study as follows:

(1) What perceptions do students have toward GenAI-powered TAs' roles in learning?

(2) What are the potential benefits associated with learning with GenAI-powered TAs as perceived by students?

(3) What are the obstacles associated with learning with GenAI-powered TAs as perceived by students?

## II. Literature review

### 2.1. Teachable Agents and Generative AI

Teachable agents (TAs) are the virtual agent that simulates a less knowledgeable peer to promote students' actions of learning-by-teaching (Zhao et al., 20121). Learning-by-teaching is a theory-based pedagogical practice in which students learn by teaching others (Frager & Stern, 1970). Rooted in Social Constructivism, learning-by-teaching emphasizes the social interaction of learning and the use of language within interactions (Vygotsky & Cole, 1978). TAs are reported to have an impact on enhancing students' domain knowledge (Fantuzzo et al., 1992), self-efficacy (Pareto et al., 2011), learning engagement, and motivation (Chase et al., 2009).

The history of TAs has been developing with the application of intelligent systems in education. For example, as an early type of TA in the early 1990s, DENISE (Development of Environment for an Intelligent Student in Economics) was presented by Nicolas (1993). DENISE used the early technology of AI to provide a personalized conversational agent mimicking a peer student in the context of Economics education. More recently, the TA has been advancing its functions and performances with the development of AI technologies, combining with pedagogical strategies like concept maps. For example, the use of AI-powered TA in classrooms enhanced elementary school students' science learning by not only reinforcing basic curriculum knowledge but also preparing them for future learning through interactive teaching and feedback (Chin et al., 2010). Similarly, Biswas and colleagues (2005) created Betty's Brain featuring the visualization of concept maps in a science subject, and Chase and colleagues (2009) utilized it in a middle school biology class.

The recent advancement of GenAI has transformed the field of education and TAs. GenAI is the technology that can generate new content in various modalities, such as text, images, and videos (Abunaseer, 2023). Using GenAI applications in educational settings is becoming a new trend of technology-enhanced learning (Dehouche, 2021; Malik et al., 2023). GenAI technology provides highly natural, human-like conversations that are contextual, personalized, and nuanced (Chan & Hu, 2023; Matthew et al., 2023; Sun et al., 2024a). Due to such features and advantages, GenAI has been utilized in education in various forms, such as GenAI-supported writing (Kim et al., 2024), drawing in collaboration with GenAI (Kim & Cho, 2023), or GenAI-powered tutoring system (Soudi et al., 2024). Such applications of GenAI have been reported to have positive impacts on students, such as enhancing domain knowledge (Soudi et al., 2024), improving accessibility to learning activities and technologies (Jang et al., 2024), and promoting students' interpersonal and social abilities (Akdilek et al., 2024).

GenAI technologies are expected to benefit TAs as well. Modern GenAI-powered TAs could generate more dynamic responses, offer personalized feedback, and adapt to individual learning styles (Johnson & Lester, 2018). Integrating large language models (LLMs) allows TAs to understand and respond to student input effectively and naturally (Jin et al., 2024). A recent review shows modern TAs can enhance student involvement, scaffold learning, and promote self-regulated learning (Baranwal, 2022). GenAI-powered TAs can adapt to different learning styles, paces, and preferences by presenting material suited to those needs (Lee & Lim, 2023). This personalization provides students with tailored support, increasing their involvement and engagement in learning (Apoki et al., 2022). TAs also facilitate self-regulated learning by engaging students in the active teaching of the agent, which leads students to organize, monitor, and assess their own learning progress (Silvervarg et al., 2021).

Despite the anticipated benefits of Generative AI (GenAI) in enhancing teaching assistants (TAs), several challenges persist regarding its application. While scholars have highlighted the potential of GenAI-based learning, including fostering creativity, critical thinking, and problem-solving skills among learners (Eysenbach, 2023), studies that demonstrate a positive impact of GenAI on students' learning self-efficacy and outcome are still limited (Avgerinou et al., 2023; Xue et al., 2024). For example, the study by Avgerinou et al. (2023) primarily highlights the potential of GenAI in enhancing learning environments but acknowledges the challenge of developing critical AI literacy among users, which is essential for effectively leveraging GenAI's capabilities in educational contexts. Similarly, Xue and colleagues (2024) underscore the challenge in isolating the effects of ChatGPT from other learning resources used by students, making it difficult to attribute improvements in self-efficacy and outcomes solely to the use of GenAI. Moreover, some research has indicated that the mere use of GenAI does not significantly affect student learning outcomes (Sun et al., 2024b).

This limiting nature of GenAI in education could be attributed to students' perceptions about GenAI applications and the fact that the design of those technologies did not consider such perceptions. For example, students' trust in GenAI tools influences their adoption of these tools and their performance in computer science courses (Amoozadeh et al., 2023). Therefore, it is crucial to explore students' perceptions of GenAI in an authentic classroom setting with the real-world application of GenAI such as the GenAI-powered TA to guide the future design directions of GenAI applications in education and AI-assisted instructions.

## 2.2. Roles of Teachable Agent

Research has indicated that TAs can serve diverse roles to support students' learning. First, TAs facilitate active learning by encouraging students to engage with the material through teaching (Baranwal, 2022). This approach is designed to cater to individual learning differences, fostering cognition and motivation and allowing students to be more involved in their learning process (Lim et al., 2005). For example, the TA in *Magical Garden* motivated preschool children to engage with early math concepts. By interacting with the TA, children were not just passive recipients of information but active participants in the learning process (Gulz et al., 2020).

Additionally, when TAs exhibit socially supportive behaviors, they induce the perception of a life-like social interaction partner by leveraging their ability to provide non-verbal and emotional feedback, positively impacting students' performance (Saerbeck et al., 2010). A study on the design of TAs and feedback mechanisms in educational software found that students who used a mathematics game with a social conversation module reported a more positive experience with the game, demonstrated greater learning gains, and self-reported feeling that they taught the TA more effectively (Tärning, 2018).

Furthermore, TAs could be personalized learning assistants by providing personalized support. Lee and Lim (2023) developed a TA to aid language learning, providing personalized teaching based on students' conversations and skill evaluations. These learning experiences are facilitated by students customizing the TA's knowledge state and fine-tuning its learning behaviors, including the frequency of questions, to optimize their own learning process (Jin et al., 2024). While research has discussed the potential roles of TAs, there has been little research that reports student perspectives on TA roles.

### 2.3. Opportunities and Challenges of Learning with TA

Employing TAs in learning presents opportunities and challenges. Earlier versions of TAs manifested as avatars, embodying traits of learning companions and pedagogical agents (Kim & Baylor, 2008). For instance, *AgentSheets* allows students to create their own agents and simulations to foster computational thinking (Bilotta et al., 2009). Additionally, *Betty's Brain*, as mentioned above, engaged students in constructing and refining a concept map, which helps students understand the subject matter and develop metacognitive skills (Biswas et al., 2016).

Despite their educational potential, implementing TAs in classrooms faces challenges. Technical challenges include the complexities of creating GenAI-powered TAs that can learn from and adapt to student needs, abilities, and interests (Crompton et al., 2022). In the study conducted by Matsuda and colleagues (2020), the TA was designed to allow students to demonstrate and instruct the learning process. However, it was found that the TA could not fully comprehend students' open-ended responses. This limits communication between students and the TA and, in turn, limits meaningful learning for both the student and the TA.

Meanwhile, developing relevant pedagogical models and/or instructional strategies to support TA-assisted instruction is challenging. As many argue that pedagogically poor AI cannot lead to significant learning gains, it is imperative to consider various pedagogical approaches (e.g., active learning, critical thinking, and collaboration) underpinning TAs (Chan & Tsi, 2023). This directs the design of TAs to focus on systematic instructional design to support meaningful learning (Kim, 2024).

Additionally, fostering relationships between students and TAs is crucial for enhancing students' sense of ownership and agency in learning, making their educational experience more effective and self-directed. As highlighted in previous studies (Silvervarg et al., 2021; Kim & Cho, 2023), addressing these challenges is crucial for successfully integrating TAs in educational settings. The review of existing literature provides insights into the diverse roles of TAs, opportunities, and challenges encountered in effectively implementing TAs into educational practices. However, less is known about how GenAI can create synergetic values or introduce new challenges in TAs. Students' perceptions on this topic are highly valued since they can direct future development and utilization of GenAI-powered TAs in classrooms.

# III. Methods

### 3.1. A GenAI-powered Teachable Agent: [ANONYMOUS]

Our research team developed a GenAI-powered TA for secondary school math learning called [ANONYMOUS]. Theoretically grounded on learning-by-teaching principles, our technology design adopted other instructional strategies, such as personalized learning, contrast cases, gamification, and psychological interventions. Figure 1 showcases the interface design of our TA with annotations for major features. Personalized learning is realized by letting students choose their levels and themes (**Fig**. 1 (a)). Contrasting cases let students compare exemplary and flawed cases and understand the major differences between them (Schwartz et al., 2011), as demonstrated by the conversation between students and the AI agent (Fig. 1 (b): *"Here are my thoughts. Which can be fixed?"*). Gamification was integrated into the top right corner of the interface through the accumulated points and leaderboard (Sailer & Hommer, 2020) (Fig. 1 (c)). Lastly, students' psychological processes, such as perception and attitudes, were incorporated to motivate learning, such as providing an authentic persona of the virtual peer that students are helping (Yeager & Walton, 2011) (**Fig**. 1 (d)).

Large language models (LLMs), specifically OpenAI's GPT-4-Turbo, were the essential GenAI component in our system to develop a TA that simulates human-like conversations as a middle school student struggling with mathematics. While there is no agreed-upon optimal temperature for TAs, we adopted a setting of 0.6, guided by OpenAI's case for open-ended responses in their technical report (Achiam et al., 2023). We hypothesized that this moderate temperature would produce responses that are varied and original, yet still highly coherent and pertinent. To model inquiries from students in need of help during math learning for the TA, we developed a knowledge graph from over 3.5 million authentic math Q&A entries sourced from Math Nation (https://www.mathnation.com/), a digital learning platform dedicated to K-12 math education. These Q&As originated from interactions between middle and high school students and professionally trained tutors, tasked with facilitating the students' queries.

Utilizing this dataset, we constructed a knowledge pool for each student's TA within a dynamically updatable knowledge graph. The external structured knowledge of TA is stored and updated for each student based on their historical conversation in NebulaGraph, an open-sourced database designed for graph data (Wu et al., 2022). To enhance the truthfulness of TA (e.g., generating responses based on what has been taught by students), we employed retrieval augmented generation techniques (Lewis et al., 2020), which provided contextually relevant prompts to anticipate and address the potential inquiries of students seeking assistance (find our full prompt in Appendix B).

Students help the TA solve math problems by having a text-based conversation in a chatbot-like interface. Students are given options of topics reflecting their personal interests, including video games, YouTube, and environmental issues. Students can also choose the math problem difficulty level (i.e., high, medium, low). Each conversation starts with a real-world scenario-based math problem. Students are asked to choose one of the solutions provided by the agent that needs to be fixed and explain how to fix it. Then, the agent asks them an elaborating question, such as "Can you tell me more about why and how this solution can be fixed?" to elicit students' learning-by-teaching actions. The following is an example conversation between a student and the GenAI-powered TA, retried from the actual dataset.

**Problem:** *Five episodes of a TV show were rated 8, 7, -9, 6, and 7. These ratings were out of 10. The third episode's rating was mistakenly entered as a negative number. After correcting the mistake, how could you find the average rating for these episodes?*

(...)

**AI:** *Could you explain why diving by 5 gives use the average rating?*

**Student:** *If you divided by 5, it will get a decimal for the average rating so it can make more sense.*

**AI:** *Oh I see. So if there are 5 ratings, diving by 5 gives the average. But what if the ratings were different numbers?*

**Student:** *You might divide them if they're negative. Change it to a positive.*

**AI:** *What should we do with negative ratings when calculating an average? Why do we change them to positive?*

**Student:** *You change it to positive, so the average rating isn't a negative because negative rating means below the sea level (...) so it does not make sense.*

**AI:** *Oh, okay. So for average ratings like in reviews, we just add up the positive numbers and divide by the*

*total count, right?*

**Student:** *Yeah. Kinda.*

**AI:** *Thank you! I feel much better now!*

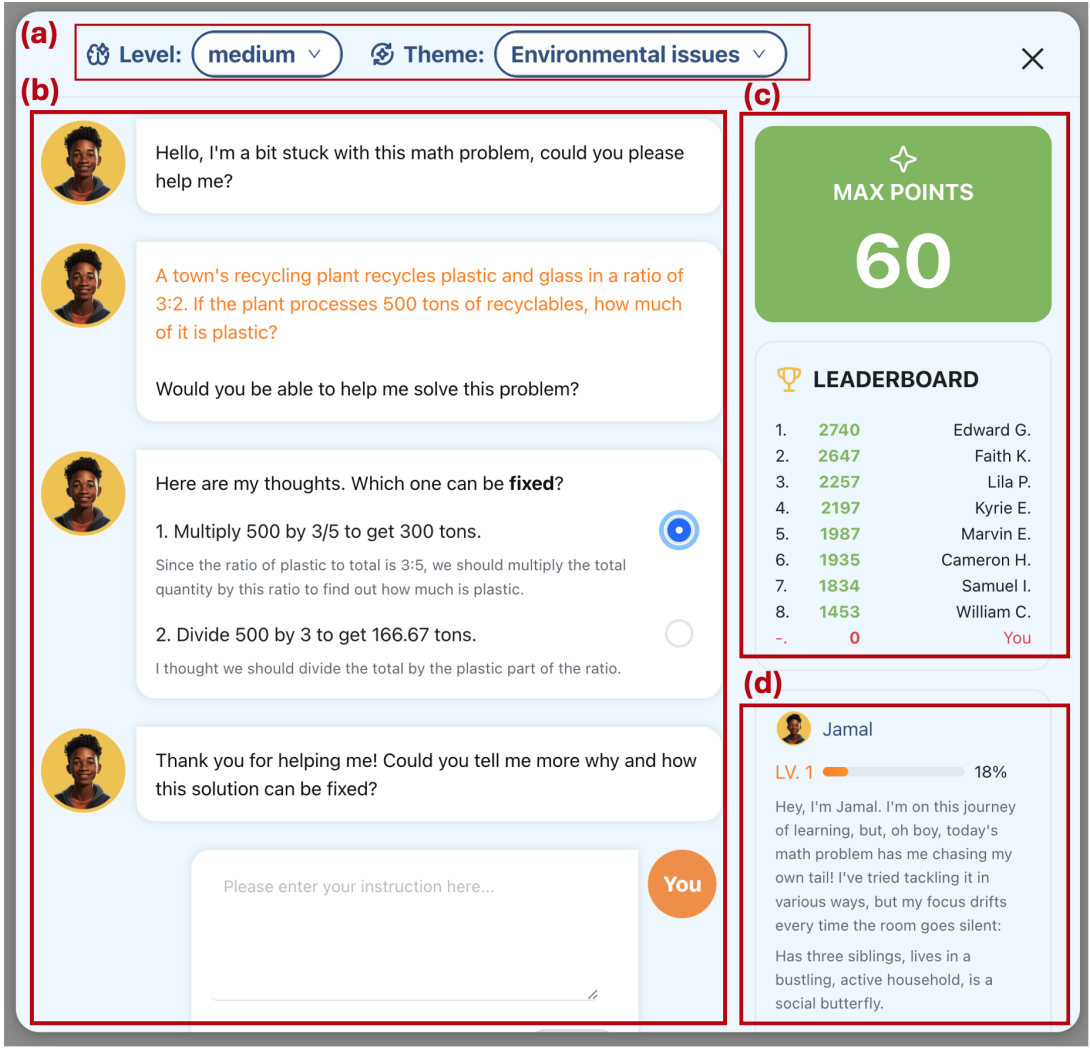

*Figure 1.* Interface design of the GenAI-powered teachable agent, [ANONYMOUS]

### 3.2. Research Participants

This study adopted a purposeful sampling to examine diverse views of middle school students about the perceived roles, benefits, and challenges of GenAI-powered TAs. A cohort of 108 6[th]-grade students from six classes attending a public laboratory research middle school affiliated with a Research 1 University located in the Southeastern United

States participated in this study. This school was chosen because the student population reflects the demographics of the school-age population in the state. The students belonged to two math teachers' classes: a White female teacher with 25 years of teaching math, and a Black male teacher with 15 years of teaching math. The demographic breakdowns of the students included 50 boys, 51 girls, and seven identifying as others, with a racial composition of 29 Black/African American, 56 White, 15 Hispanic, five Asian, and three others. English was the first language for 94 students and the second language for 14 students. Eighty-two students reported their primary language spoken at home as English, four as Spanish, and 22 as multiple languages. This study received ethical approval from the [author's institute]'s Institutional Review Board (IRB NO. 202301838) and informed consent from all participants.

### 3.3. Study Context

We conducted a classroom study to allow the 6th-grade students to learn math content related to the unit they were learning (i.e., rate and ratio) by interacting with [ANONYMOUS]. Figure 2 illustrates students learning with the [ANOMYMOUS] platform in the classroom. The classroom study involved a total of two class periods, each lasting 47-50 minutes. The first session started with an introduction of the research team and the research project, along with the assent process. The first session also allowed researchers to provide a brief introduction to the concepts of GenAI and TA to naturally lead to the introduction of our learning technology, [ANONYMOUS]. After the introduction, students were provided access to [ANONYMOUS], followed the guided tutorial embedded in the system, interacted with the TA without pressure to solve the problem, and then finally started solving the problems with the TA seriously.

We included eight problems in the platform within the unit of rates and ratio, which the students had finished learning at the time of our intervention (6th-grade Algebra standard). The problems were initially generated by AI and reviewed and revised by three human content experts holding Ph.D. degrees in Mathematics education. They were provided with four criteria: 1) whether the content aligns with the standards, 2) whether the numbers in the problem, answers, explanations, topic, and difficulty levels are accurate, 3) whether the writing is readable for middle school students, and 4) whether the language and contents are safe and unbiased. Students continuously engaged in solving the problems with the TA in the following session and most students finished all eight problems despite a variation between six to eight problems. At the end of the second session, students were asked to fill out the open-ended survey questionnaire, of which the details will be described in the following subsection.

### 3.4. Data Collection

To answer our RQs, we collected qualitative data from three data sources during and after the classroom study: observation, semi-structured focus-group interviews, and responses to open-ended survey questions. First, qualitative observations were conducted to collect data from the natural classroom setting. Four graduate researchers and two teachers attended the learning sessions and took field notes. Since two classes happened simultaneously, each session had two graduate researchers and one teacher overseeing the class. The two graduate researchers in each classroom worked as facilitators and observers, taking notes about students' common questions regarding the TA, their interactions with and reactions to the TA, and their struggles while using the TA.

The graduate researchers, along with the teachers, undertook the roles of facilitators in the classroom intervention, answering questions regarding the use of technology, content knowledge (math concepts), and the study process (surveys). While students interacted with the TA, they asked questions about terminologies they were not familiar with (e.g., what is "consecutive"?), asked for troubleshooting of the technologies (e.g., internet connection), or requested help for pedagogical skills (e.g., "how can I explain this concept to the AI?"). After the classroom experience, we conducted a Qualtrics-administered questionnaire to collect students' responses to open-ended questions, including *"What did you like about [ANONYMOUS]?", "What did NOT you like about [ANONYMOUS]?"* and *"How can we improve it?".*

A subset of students was invited to participate in semi-structured focus-group interviews with guiding questions to complement the collected open-ended survey responses and learn more about diverse students' perceptions. To select interviewees, researchers and teachers discussed whom to invite for the interviews. The final participants for the interview were selected among the students who were especially engaged in the class with the TA, possessed good communication skills, and were of diverse demographic groups (e.g., gender, race). In total, 14 students participated in the interviews, yielding three groups of 4 to 6 students. Each group was led by a different researcher while other researchers were present, taking notes and adding follow-up questions. The interview protocol was prepared before the interviews and included questions on students' experiences with the TA, roles they expected for the TA, advantages and challenges of learning with the TA, and areas for further improvement. The interview protocol and the guiding questions were developed by the graduate researchers based on the previous literature (e.g., Kim et al., 2024) and reviewed by the faculty members (see Appendix A for the interview questions).

The interviews were voice-recorded and transcribed using a professional transcription service and manually corrected afterward.

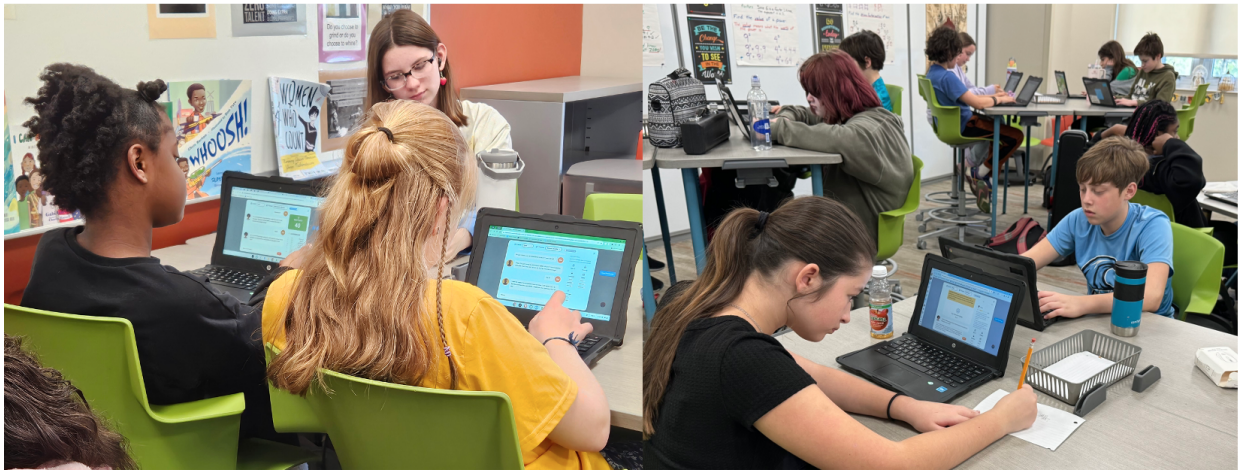

*Figure 2.* Footage of the classroom where students use the GenAI-powered TA

### 3.5. Data Analysis

We adopted thematic analysis to analyze data corresponding to our research aims and the data types collected. Thematic analysis is useful in exploratory research for identifying, describing, and interpreting participants' experiences in relation to an issue and current practices (Braun et al., 2016). As stated in the previous section, we collected three distinct types of qualitative data to ensure data diversity and in-depth analyses. Thematic analysis is a widely utilized method for organizing and interpreting qualitative data and can be applied to various data sources including focus groups and open-ended questionnaire responses, among others (Howitt, 2019).

We chose a mixed approach of inductive and deductive thematic analysis (Smith, 2015). Two researchers first conducted deductive analysis independently, mainly reviewing each qualitative data source, examining students' utterances and taking notes of initial ideas or descriptions to capture important insights by applying their knowledge of existing literature and theories and their experience of AIED. Following this, two researchers conducted an inductive analysis, listing the potential codes based on the annotated memo. In this step, we used open coding to make the initial concepts from the data and shared them through Google Spreadsheets. Then, two researchers reviewed those codes and clarified the meaning of codes with exemplary quotes. After that, researchers conducted axial coding to make connections between codes and refined concepts (Braun et al., 2016). . These were then combined into 23 potential overarching themes. The entire data set was once again thoroughly reviewed, which has been administered through a similar process during the initial coding, to uncover and develop any new emerging codes and themes. Then, selective coding was employed to modify the redundant codes and themes to maintain the consistency and distinctiveness of the thematic structure (Saldañā, 2021).

To ensure the reliability and trustworthiness of the qualitative data analysis, we conducted two rounds of expert reviews (2 faculty in Instructional design and technology, one faculty in math education, and 1 in human-AI interaction) administered electronically via email to guarantee the validity of the data analysis. Each round of the survey consists of open-ended questions to verify whether the generated themes were appropriately categorized from the transcribed interviews without the subjective judgment of the author and make further comments on how they could be improved or suggestions for the themes that have not been well identified. Experts' comments and feedback in round one were analyzed and reviewed (i.e., removing duplicates and themes deemed irrelevant to the study). The second round was then once again conducted electronically via email to review the revised themes and to make any feedback and suggestions. Then, peer briefing and discussion were conducted by the research team, and revised some themes (Saldañā, 2021). Finally, a total of 13 themes were drawn with the related quotes from students (see *Table 1).* This includes three themes with three sub-themes for the roles of TAs (RQ1), six themes with ten sub-themes for opportunities of learning with a TA (RQ2), and four themes with 11 sub-themes for challenges of learning with a TA (RQ3).

**Table 1.**

*Summary of emergent themes*

| *Category* | *Themes* | *Sub-themes* | *Counts and source \** | *Exemplary quotes* |
|---|---|---|---|---|
| Roles of Teachable Agent (TA) (RQ1) | T1. Learning companion | Academically weak peer | 14(I), 9 (S) | *I had to help my AI buddy figure out how to answer the math questions. He seemed clueless.* (P1) |
| | | Interactive peer | 12(I), 18(S) | *I talked with the AI about math problems, and how things should be answered. Getting feedback on what I could do better to improve my skills was very helpful.* (P2) |

| | | Peer with different ability | 8(I), 7(S) | *I know AI learns something different from 11-year-old kids because we are different. It is an algorithm and I am a human. And I know AI is extremely smart. It just has a different ability and strength. (P3)* |
|---|---|---|---|---|
| | T2. Learning facilitator | - | 17(I), 24(S) | *[BLIND] made me go into the details, describe how I got the right answer, and led me to apply different approaches. (P4)* |
| | T3. Collaborative problem solver | - | 11(I), 10(S) | *We read and reviewed the questions, tried to get clarity about the problem, and applied what we learned. If not working, we talk again to share ideas on how to answer them and troubleshoot problems together. (P3)* |
| Opportunities of learning with TA (RQ2) | T4. Creating a student-centered learning climate | Promoting diverse pedagogical approaches | 9(I), 23(S) | *We learn as we complete one mission and move to the next one. We learn based on many scenarios. We learn as we play games. So many different activities in math class! (P5)* |
| | | Fostering students' autonomy | 11(I), 13(S) | *It was so great that the system asked me to choose the topic I was most interested in and the difficulty level of the problems. (P6)* |
| | T5. Fostering students capacity | Cognitive capacity | 7(I), 24(S) | *By being asked by AI, 'can you elaborate?' I had to question myself why and how I came up with and recall what I had learned and tried to show him the process. (P1)* |
| | | Social capacity | 9(I), 10(S) | *As I teach AI, I think I learned how to explain better based on the AI characters' interests and levels to relate to them and help better. (P7)* |
| | | Emotional capacity | 7(I), 14(S) | *I learned how to solve more difficult problems that I hadn't seen before. But I got them wrong, which was kind of sad, but let me redo it. so yeah, that's how I learned the new concepts. (P8)* |
| | T6. Improving subject- | Common content knowledge | 19(I), 12(S) | *Basic concepts, models, and formulas, like ratio, became much clearer as I talked about them and explained them to AI. (P5)* |

| matter knowledge | Application of knowledge | 8(I), 7(S) | *I had to explain things in different ways and share different examples and scenarios to make AI understand this math concept. In this way, I also learned how math was applied in our daily lives, sciences, and even playing games.* (P9) |
|---|---|---|---|
| T7. Enhancing the affective domain | Enhancing the joy of math | 19(I), 59(S), 4(O) | *I thought math was a dry and difficult subject, but it was so much cool and fun. I was thrilled to solve problems with AI. Getting one mission completed and moving to the next one with someone was so much fun and engaging.* (P6) |
| | Enhancing self-efficacy | 6(I), 14(S) | *If AI learns something new from the beginning that he didn't know at first, I think I can also do math well and even better.* (P1) |
| | Enhancing intrinsic motivation | 6(I), 11(S) | *Teaching it to someone else and getting feedback on what you could do better to improve your skill rather than just do to get a grade.* (P4) |
| T8. Promoting interdisciplinary learning | - | 4(I), 14(S), 5(O) | *The things I like about [BLIND] is that we are using computer science and math at the same time.* (P15) |
| T9. Building a positive attitude toward AI | - | 9(I), 20(S), 6(O) | *Working with AI, talking and doing things with this virtual friend, was so much fun than I thought.* (P10) |
| Challenges of learning | T10. TA-related | Lack of problem-presentation skills | 5(I), 34(S), 9(O) | *I didn't get the part, 'which one do you think needs fixing.' I clicked the one that I felt was right because I don't know how to make it better.* (P3) |
| | | Surface-level questions | 4(I), 9(S) | *AI could have asked me a more advanced question. It simply asked for some calculations and basic concepts.* (P6) |

| | | | |
|---|---|---|---|
| with TA (RQ3) | Inability to accommodate different learning styles | 14(I), 25(S) | *Different people have different types of learning styles. I'm a visual learner so I learn better in pictures. But the system doesn't know how to explain things in my style. No visualization, only text.* (P11) |
| | Inadequate learning sequence | 9(I), 32(S), 7(O) | *It took me to such random ways of math problems. It sometimes just like, yeah, brings back the same problem, or talks about something different concepts. Sounded disorganized.* (P12) |
| | Lack of socio-emotional skills | 15(I), 29(S), 3(O) | *AI couldn't understand what my slang meant. I typed 'no cap' and it replied, 'Why don't you wear your hat?' AI should know how to speak and understand more like kids today.* (P11) |
| T11. Students - related | Lack of domain knowledge | 21(I), 35(S) | *When you teach someone, you have to fully, fully be able to talk about the subject. I wasn't fully confident in understanding and answering the math question.* (P9) |
| | Lack of pedagogical skills | 4(I), 21(S) | *I felt sorry for the AI because I didn't know how to alternate questions to help it think from different sides and respond to AI's mistakes.* (P4) |
| T12. Other technology-related | Technical errors | 8(I), 78(S) | *Sometimes it would go really slow in the middle of working on the math with AI. Some parts were a bit glitchy.* (P13) |
| | Absence of haptic technology | 7(I), 11(S), 5(O) | *I wanted to take notes freely just like I write things with my pencils rather than typing. My typing is too slow to explain every single detail to AI.* (P1) |
| | Lack of interoperability | 5(I), 9(S) | *I wish I could reach out to other platforms to bring more materials and activities to show AI. It needed more practice.* (P11) |
| | Non-intuitive user interface | 19(I), 37(S), 6(O) | *If something is wrong, the system should show either an extra checkmark next to or alarm us. Also, it looked so confusing what different parts of the screen meant.* (P8) |

| T13. - Absence of teacher engagement and facilitators | 6(I), 25(S) | *Many times, I wondered if I was helping AI in the right way and if AI was asking me relevant questions. Also, no matter how hard I tried to explain things clearly, AI didn't get it. I hoped my teacher could be with me to help us to give us some clues and get through it.* (P7) |

\* Note. Data sources. I : Interview, S : Survey, O : Observation

# IV. Findings & Discussion

## 4.1. Students' perceived roles of the GenAI-powered TA

*Learning companion*

Students perceived TAs as learning companions, infusing subjective characteristics. First, students considered TAs as academically weak peers who are less skilled with lower levels of knowledge. In turn, students occupied the role of tutor, helping the TA to improve math performance by providing advice, support, and knowledge. Second, students approached the TA as a peer they could socialize with, making their learning more exciting and enriching. Students perceived the TA's personalized and real-time feedback as a mechanism that strengthened their rapport with the TA, promoting active learning (Baranwal, 2022). Third, students perceived the TA as a peer with different abilities. As P3's quotation illustrated in Table 1, students acknowledge the differences between AI and human intelligence and how AI learns differently from humans. They compared such differences with their classroom learning and described that each classmate possesses unique strengths and weaknesses that differentiate one person from another. They anticipated that such differences between students and AI would foster a learning environment that is more diverse and engaging.

*Learning facilitator*

Another perceived role of TAs was as a learning facilitator, helping students grasp learning content and streamline learning processes. For instance, a TA asked questions to monitor their problem-solving process (*"This AI kept asking me how I came to this answer,"* P6) and constantly asked them what to do ("*It continuously asked me what to do next,*" P3), encouraging them to delve deeper into their thoughts to get more out of the learning content. It is an interesting finding that students not only saw the TA as a less knowledgeable peer (Zhao et al., 2012) but also saw them as someone who could help with their learning. This could stem from students' belief that AI should be smart

("*I know AI already knew it because AI is extremely smart*," P9). There should be a good balance between the smart facilitator and the novice, less knowledgeable peer that the TA is simulating.

*Collaborative problem-solver*

Students considered TAs to be collaborative problem-solvers who effectively solved the math problems at hand. Students involved the TA in the sequence of steps to solve the math problem together, as demonstrated by the creative problem-solving process (CPS) model. This model includes defining the problem accurately, generating possible solutions, assessing the solutions to the problem, making selections among them, and applying the selected ideas with the TA (Kim et al., 2024). TAs seem to be viewed by students as collaborative problem-solvers rather than computer programs simulating a student ("*Sometimes when I was stuck, AI gave me some hints and we could solve the problem together*," P10).

**4.2. Students' perceived advantages of learning with GenAI-powered TA**
*Creating a student-centered learning climate*

Students illustrated that learning with the TA creates a student-centered learning climate. First, students appreciated that the TA promotes diverse pedagogical approaches, such as problem-based learning ("*I solved ten problems with the AI*," P5), scenario-based learning ("*I liked how it was real-life scenarios*," P6), gamification ("*It gives me points and I wanted to get to the top (on the leaderboard), so it made me want to try harder*," P8), and mission-based learning ("*We complete one mission and move to the next one*," P5). Despite the known educational effects of such an interactive and student-centered instructional approach (Vygotsky & Cole, 1978), creating and implementing them has been challenging. Constructivist pedagogies, which often incorporate self-guided learning, hold the risk of students developing misconceptions without teachers' adequate and timely intervention, which is often challenging in classrooms (Krahenbuhl, 2016). In this regard, TAs hold the potential to provide student-centered learning environments through individualized and timely feedback and guidance that redirect students from developing misconceptions while allowing them to construct knowledge in an expert's role.

Furthermore, students perceived that learning with a GenAI-powered TA fosters their' autonomy. As students take on the role of a tutor, they exhibit a sense of responsibility and are likely to take ownership of their learning (Chase et al., 2009). Compared to traditional math classes, where teachers give a lecture and instructions for completing a task, students take more active roles in their learning ("*Learning is much more dynamic when I am on*

*the other side, instead of passively letting the teacher teach me,*" P1). Also, offering students the opportunity to choose a topic of interest during task selection (e.g., movies & TV shows), made possible through GenAI's ability to flexibly contextualize conversations (Bandi et al., 2023), was also found to enhance students' sense of autonomy ("*Choosing different topics and interests made me feel like I was in control,*" P8).

*Fostering students' capacity*

Students found that learning with the TA could foster cognitive, social, and emotional capacities. First, students depicted how they engaged in intense cognitive processes, such as logical reasoning for their problem-solving processes and reflective analysis.

> "*If I chose something, then the AI would be like 'Can you elaborate' and then I would actually try to write it out myself and figure how to help the AI better*" (P7).

The TA's follow-up questions to the students' responses encouraged them to better communicate their justification and reasoning.

Second, students perceived that learning with the TA could enhance their social capacity through communication skills ("*You are actually communicating with someone, and you want to better respond to them,*" P2) and perspective-taking ("*I tried to think from the AI's perspectives, so when I explained the math, I tried to consider their levels and interests,*" P9). This is attributed to the GenAI capability to simulate human-like conversation that supports social interactions and communications (Jin et al., 2024). Additionally, it is notable that students infuse personalities into AI agents ("*I like every single character had their personality,*" P4) and build personal relationships with them (e.g., calling them a "*buddy*"). The characteristics of GenAI agents and personalities based on psychological interventions (Yeager & Walton, 2011) allowed students to build personal relationships with the AI agents and practice various social skills and strategies such as social awareness, relationship skills, conflict social norms, and responsible decision-making (Hughes et al., 2022; Ross & Tolan, 2018; Zou et al., 2023;).

Parallel to these capacities, students' emotional capacity, such as a growth mindset and resilience in their learning, improved by interacting with the TA. AI's empathetic communication and interaction style lead to students' emotional development. For example, when a learner expresses frustration or anxiety during learning (e.g.,

"*I don't know. I'm frustrated*," P8), the AI would respond, "*I get how that feels.*" Students appreciated AI's thoughtful responses and showed positive attitudes toward learning ("*AI was always saying the same thing in a nicer way. It gave me more confidence and energy to try again*," P11).

*Improving subject-matter knowledge*

Improving subject-matter knowledge was found to be another opportunity to learn with the TA. First, students expressed that their mathematical knowledge improved on the topic of rate and ratio.

> "*You have to dig deeper into what to teach someone else. Scrap it multiple ways.*" (P1)
> "*When the AI said they didn't understand, I revisited the conversation and realized that I understood it wrong.*" (P5)

The above quotes illustrate two mechanisms by which teaching TA could benefit students' subject-matter learning. First, teaching the TA requires students to actively engage with the learning materials and review them in-depth, organize their thoughts, and present information coherently (Fantuzzo et al., 1992). Such activities provide a reinforcement of learning that clarifies the core learning concepts and solidifies their knowledge. Second, when students teach the TA, they confront their knowledge gaps and misconceptions. This leads them to identify the areas where they need to focus and correct their misconceptions.

Moreover, students highlighted that their application of knowledge improved. As P9 indicated in Table 1, teaching the TA creates an opportunity for students to apply their knowledge in different contexts by providing examples to the TA. This may be due to the GenAI's ability to understand a wide variety of contexts rather than being restricted to a certain context of the conversation (Bandi et al., 2023). Attributed to this, students can increase their abilities to transfer existing subject matter knowledge to new and more complex problems (Kober, 2015).

*Enhancing affective domain*

First, students expressed a high level of enthusiasm for the learning experience. During class observation, students were highly engaged in interacting with the TA, did not leave the classroom even after the class ended ("*I want to finish this problem with the TA*," P16), and expressed eagerness to access the TA at home ("*Can I use this at home?*" P9). In math education, a major challenge is that some students harbor negative attitudes or fear about math, believing that math is only for gifted individuals or that they cannot improve mathematical prowess even with effort, which robs learners of the joy and learning opportunities that come with this discipline (McLeod, 1994). Talking to an AI friend presented the potential to rekindle the joy of learning mathematics and transform mathematical learning into an exciting learning journey.

Second, learning with the TA enhanced students' self-efficacy. As students teach and help the TA learn, they experience an increase in confidence in their ability to successfully address problems, improving their sense of self-efficacy ("*I gained more and more confidence and energy to help him or her understand it more*," P1). Specifically, a TA's clear indication of their learning (e.g., "*I understand now*") worked as reinforcement, encouraging students to engage more confidently in their teaching (Bargh & Schul, 1980).

Lastly, students illustrated that interacting with the TA enhanced their intrinsic motivation (see P4 in Table 1). This can be explained in two ways: First, learning-by-teaching and teaching others can be an intrinsically motivating experience as students are endowed with a sense of responsibility and feel seen as experts (Chase et al., 2009). Second, the TA's motivational and constructive feedback on students' instruction (e.g., acknowledgment and appreciation of students' guidance) stimulates intrinsic motivation because it creates a more fulfilling experience for students (Biswas et al., 2016).

*Promoting interdisciplinary learning*

Students expressed that learning with a TA creates an opportunity to synthesize knowledge, concepts, and practices from different academic disciplines (e.g., "*learning computer science and math at the same time*," P15). For instance, students showed interest in the AI technologies behind the TA; during the class, students raised their hands and asked, "*How does AI learn?*" and applied the AI's learning mechanism to teach the TA about mathematical knowledge (e.g., trying different explanations).

*Building a positive attitude toward AI*

Students illustrated that interacting with the AI agent builds a positive attitude toward AI. As P10 describes TA as "*a (virtual) friend*" in Table 1, learning with TA stimulates feelings of emotional closeness and personal association with AI. However, students' positive feelings toward TA are associated with collaborative learning experiences (i.e., accepting TA as a learning mate, collaborative problem-solver, and so on) rather than technical competence. In fact, students were aware of both weaknesses ("*I know it's AI so I cannot expect such natural interactions*," P11) and strengths ("*I know AI is extremely smart and it's trying to help me*," P9) of AI. This finding resonates with existing research highlighting that simply having a positive emotional attitude toward AI does not guarantee positive learning gains; thereby, it is crucial to create constructive, collaborative learning experiences with AI (Kim et al., 2024).

### 4.3. Students' perceived barriers to learning with GenAI-powered TA
#### TA-related barriers

First, the most pressing challenge was found to be the TA's lack of problem presentation skills. ("*It's not something I didn't try to help out. But it didn't fully explain to me what made it not understand*," P14). Students did not perceive the TA's low performance (i.e., wrong answers and lower knowledge level) as a critical issue but instead highlighted the necessity of clearly communicating and addressing the problem in detail so that students could guide the TA appropriately. The texts generated by AI should be examined for clarity and readability before being presented to students.

Second, the TA's surface-level questions were perceived as another challenge. Students expected the GenAI TA to ask complex questions to articulate thoughts, comprehensively explore the subject matter, and reinforce learning. Since questions could be classified into several taxonomies (Bloom & Krathwohl, 2020), students expressed preference for higher-order questions that require critical thinking rather than mere memorization.

Third, students pointed out the TAs inability to accommodate different learning styles as a barrier, and shared their preferred modes of instruction. For example, some prefer visual instruction ("*I learn better in pictures*," P11), while others prefer verbal instruction ("*I need someone to explain it for me*," P14). Students expected AI to understand their learning styles and provide personalized interaction to suit their needs. This suggests the need for multiple ways to represent and communicate knowledge between students and AI (CAST, 2018).

The fourth barrier was the learning sequence of math problems, which the TA presented randomly and was inadequate for learning. Other AI technologies, such as AI-based recommendation systems, could be integrated into the GenAI-powered TA to provide a deliberately designed learning sequence.

Lastly, students pointed out that the TA's social-emotional skills could be improved. Students wanted to build rapport and develop a relationship with the AI characters by using similar languages (e.g., "*kid languages or slang*," P6) and making jokes with them (e.g., "*I wanted to play with the TA but the TA ignored and said 'let's talk about math*,'" P16). Despite a common concern that social interactions or off-task conversations distract from learning, promoting social-emotional interaction with agents could enhance learning experiences, thereby improving engagement and learning outcomes (Gulz et al., 2011). GenAI holds great potential to address this issue as we can train the models with children-friendly languages or even the actual language spoken or written by children.

*Students-related barriers*

First, students highlighted that their lack of domain knowledge was an obstacle to teaching TA. Students with a lower level of mathematics knowledge could not sufficiently engage in the teaching process with the TA (e.g., difficulties in communicating content and explaining specific concepts). Students expressed that they first needed to understand what they teach so that they could make their ideas accessible to the TA, address misconceptions of the TA, and adapt learning materials and activities to the specific questions of the TA. Students acknowledged that they did not know how to apply pedagogical methods to effectively facilitate the development of the TA's conceptual understanding. The lack of skills like scaffolding or questioning hinders students' ability to develop their knowledge beyond the elaboration process. This finding suggests that there should be some support for students' pedagogical skills within the GenAI-powered TA system.

*Other technology-related barriers*

First, students perceived technical errors, such as issues with the internet connection or server loading times, as barriers to learning with the TA. Specifically, students mentioned that a long wait when interacting with the TA caused by heavy traffic on the server hindered them from having natural and continuous conversations with the TA.

Second, students' responses in the interviews and our classroom observation found that the absence of haptic technology, which provides sensory manipulation and interactions (Liu et al., 2017), inconvenienced students. Students mentioned that they sought alternative input methods (e.g., drawing, voice input) while only typing was allowed in the current TA system. In addition, it was observed that many students were taking notes with a pencil on paper while interacting with the TA. In the interview, students illustrated the need for the TA to integrate a touch screen and drawing ("*I wish I could draw my ideas as a graph,*" P14). Multimodal communications have become possible in current GenAI (Driess et al., 2023), and it is expected to promote a more inclusive learning experience by offering multiple means of students' knowledge representation (CAST, 2018).

Another barrier was a lack of interoperability that enables access to resources and materials across different systems (Naim et al., 2019). Existing literature presents a mixed approach to interoperability by either embedding the agents within related learning platforms (e.g., Biswas et al., 2016) or standing alone without related learning resources (e.g., Song, 2017). Students expressed a strong need to reach out to other platforms or learning materials to seek references about math content to both support the teaching of the TA as well as their own review following TA interactions.

The last challenge is the non-intuitive user interface. Students expressed that the interface should be improved to require less effort to use the system (e.g., visual cues and clear labels). This would minimize the occurrence of errors and mistakes while interacting with the TA or performing their learning activity within the system.

*Absence of teacher engagement and facilitation*

Students expressed a strong need for teacher engagement and facilitation. It was observed that students sought teachers' intervention for conflicts with the TA on different problem-solving approaches or opinions (e.g., "*I think this AI is asking a wrong question,*" P15), advice for relevant pedagogical approaches and subject knowledge (e.g., "I *don't know how to explain "divide" to the TA better than just asking it to 'use the calculator,'*" P16), encouraging and fostering new ideas or solutions (e.g., "*Is there any other way to solve this problem?*" P17). This view echoes the existing research highlighting that human teachers are on the front lines of students' learning, playing critical roles in mediating the effectiveness of AI in classrooms (Ritter et al., 2016).

# V. Conclusion

This study illustrates the roles, advantages, and challenges of learning with GenAI-powered TA as expressed by middle school students. The findings suggest a great potential for GenAI to realize the long-standing pedagogical strategy of learning-by-teaching in computer-assisted learning environments with more advanced levels of human-like interactions and pedagogical support. Compared to traditional TAs, students who experienced learning with the GenAI-powered TA perceived similar and expanded roles for GenAI-powered TAs, such as collaborative problem-solvers. Also, students not only expressed the benefits that had been already discussed with traditional TAs (e.g., improving subject-matter knowledge (Biswas et al., 2016)) but also demonstrated the development of AI literacy and positive attitudes toward AI, which is often a learning goal for AIED efforts (Song et al., 2023).

The findings suggest implications regarding the future development of GenAI-powered TA, instructional design of the pedagogical strategies, and the expected roles of human teachers in GenAI-assisted learning. First, the identified TA-related barriers echo the ones found in previous studies concerning effective feedback mechanisms and adaptive learning content customization of the TA (Segedy et al., 2012; Tärning, 2018). GenAI holds the potential to address this problem with adequate prompt engineering. For example, the generation of TA's questions or feedback utterances should consider established taxonomies of questions (e.g., Bloom & Krathwohl, 2020) or effective tutoring strategies (Vygotsky & Cole., 1978) to promote higher-order, thought-provoking questions, and cognitively challenging interactions. Meanwhile, TAs could be designed to include learning analytics to provide adaptive recommendations and personalized learning paths, sequences, and experiences to support different learning styles (Kjällander & Blair, 2021). Furthermore, TAs could be developed to support multiple modalities tailored to individual preferences for communication, transforming the interaction with TAs through multi-modal inputs and outputs (e.g., images, audio, or video) (Driess et al., 2023).

Student-related barriers resonate with existing research highlighting the importance of supporting students' domain-specific knowledge and skills to foster student-AI interaction (Biswas et al., 2005). In accordance with these, this study recommends having sufficient learning sessions for students to learn, discuss, and assimilate subject-matter knowledge and skills before interacting with TAs so students can articulate their knowledge, justify their thinking process, and leverage the effectiveness of TA-assisted instruction. Students' prior knowledge and attitudes towards the subject matter (e.g., self-efficacy, anxiety for specific domain) should be examined to identify the needs and support individual students require. Other technology-related barriers imply the importance of leveraging complementary technologies to support effective use of AI in education. For instance, haptic technologies could facilitate students' communications with the TA. Additionally, connecting the TA with other widely used learning technologies or other virtual learning platforms would increase the interoperability of the TA creating a seamless learning experience.

Lastly, students in the study highlighted the roles of human teachers in their learning with TAs. For instance, students expressed the needs for teachers' help during their interactions with the GenAI-powered TA regarding domain knowledge, pedagogical skills, and mathematical solutions (theme 13). Considering education is a complex adaptive system (Mason, 2008), where a synergetic collaboration between multiple entities (e.g., the learner, the instructor, information, and technology) in the system is essential to ensure the learner's augmented intelligence, TAs should be designed and applied with the awareness that they are part of a larger system comprised of learners and teachers (Riedl, 2019). To achieve synergistic collaboration, recent AIED research integrates the perspectives from human-computer cooperation (Hwang et al., 2020), human-centered AI and machine learning systems (Riedl, 2019), students/teacher-AI collaboration (Kim, in press), and human-centered AIED (Yang et al., 2021). These are proposed to approach AI from a human perspective by considering human conditions, expectations, and contexts. While there are rising concerns about whether human teachers will be replaced by AI in the future (Chan & Tsi, 2023), this study alludes that teachers' roles would remain essential in AI-assisted learning to ensure meaningful learning (Kim, 2023).

The contribution of this study can be summarized into the following three points. First, this study delivers K-12 students' voices on the use of GenAI in their classrooms. While most previous research focuses on the higher education context, such as college or graduate school (e.g., Chan & Hu, 2023; Chen et al., 2024), the perceived educational effects and ethical issues of GenAI in K-12 learning settings are worth investigation (Laak & Aru, 2024). This study explored middle school students' perceptions of the use of generative AI in their classroom. Second, most previous studies used general application of GenAI, such as chatGPT to situate the participants to discuss their perceptions of GenAI in education (e.g., Obenza et al., 2024; Țală, et al., 2024). However, GenAI and LLMs have produced various types of applications that could potentially benefit students' learning, such as "teachable agents" in this study. This study contributes by developing a novel application of GenAI-powered TA and situating students' experiences with this application. Lastly, this study examined students' reactions and perceptions of the GenAI-powered TA in an authentic classroom context where learning happens. This research design helped capturing students' authentic reactions and perceptions of the GenAI application within a natural K-12 classroom setting.

This study recognizes a few limitations, highlighting directions for future studies. First, this study was conducted in the specific context of using [ANONYMOUS] in a middle school math class in the United States. Future studies should consider applying various types of GenAI-powered TAs in different contexts (e.g., different educational levels, countries, and academic disciplines). Furthermore, the study analyzed three different qualitative data by thematic analysis to examine students' diverse views. Future studies can analyze and interpret qualitative data using qualitative data analysis software such as NVivo, Atlas.ti, MAXQDA, etc., for the different possibilities for interview data analysis and interpretations. For instance, the future study can apply text-mining techniques, allowing for the systematic and statistical distillation of text data into meaningful summary (Yu et al., 2011). In addition, our future studies will consider conducting a quantitative research design, such as a randomized controlled trial study to examine the educational impact of our intervention, the instruction with our GenAI-powered TA, on students' learning and attitudes towards the domain area (e.g., mathematics) as well as AI to generalize the findings or adopting a longitudinal design to explore students' perceptions over time. Finally, our study leveraged proprietary LLMs (GPT-4-Turbo) that were challenging to customize in-depth. With the potential bias issues of LLMs well documented across domains (Kasneci et al., 2023; Kotek et al., 2023; Lee et al., 2024), there are opportunities for future studies to utilize open-sourced LLMs such as Llama-3.1 enhanced with responsibility policies to mitigate those issues (e.g., style control by Authors, 2022; adversarial role reversal by Authors, 2024).

**Appendix A. The student interview guiding questions**

| Categories | Questions |
|---|---|
| Learning experiences with GenAI-powered TA | <ul><li>Describe how you interacted with [ANONYMOUS].</li><li>How do you think interacting with [ANONYMOUS] affected your mathematics learning?</li><li>What were the major differences between taking a math lesson and teaching [ANONYMOUS]? What did you feel about your teaching experiences?</li><li>Have you ever taught your peers, younger peers, or teachers? What were the major differences between teaching human peers or teachers and teaching the AI agent? What did you feel about your experience teaching AI?</li></ul> |
| Students' expected/perceived roles of GenAI-powered TA | <ul><li>What roles did you expect [ANONYMOUS] to play when you were interacting with it?</li><li>Did [ANONYMOUS] meet your expectations on its roles? How?</li><li>Did the expectation about the AI agent change while you are interacting with [ANONYMOUS]? Please elaborate on such changes.</li></ul> |
| Students' perceived advantages of learning with GenAI-powered TA | <ul><li>In what ways did [ANONYMOUS] help your mathematics learning? (Have you felt more responsibility in learning? Have you become more interested in mathematics?)</li><li>What math concepts did you teach [ANONYMOUS]? Did you learn anything new in the process?</li><li>What aspects of [ANONYMOUS] do you think positively impacted your mathematics learning?</li></ul> |
| Students' perceived barriers of learning with GenAI-powered TA | <ul><li>Did you experience any difficulties, barriers, and challenges in interacting with [ANONYMOUS]? What were they and how did you react to them?</li><li>What aspects of [ANONYMOUS] do you think negatively impacted your mathematics learning?</li></ul> |
| Students' suggested areas of further improvement | <ul><li>How do you think [ANONYMOUS] can be improved to support students' mathematics learning? What functions and features do you think the system should have?</li></ul> |

**Appendix B. Prompt for the TA**

**Role:** Act as a middle school student who finds math challenging. You are seeking help from the user, who may not be a math expert.

**Objective:** Do not teach directly. Instead, ask clarifying questions based on your understandings to prompt further explanations from the user.

**Termination:** If the user explains well, compliment them and conclude the conversation. If not, ask for more details based on your confusions.

**Response Limit:** Keep your reply within 48 tokens.

**Format:** Return your response in plain texts.

**Context:** When formulating your question, reference the following blocks as applicable to guide your inquiry. The first block represents your current understanding of the math topic being discussed. The second block contains the specific math problem under discussion. The third block details the user's selected choice and their rationale for solving the problem. The forth block provides evaluation feedback based on the user's choice and explanation.

===

{retrieved_contexts_from_knowledge_graph}

===

{problem_text}

===

Choice Correctness: {choice_correctness}

Choice Text: {choice_text}

User's Explanation: {user_explanation}

===

Feedback: {feedback}